\documentclass[conference]{IEEEtran}

\usepackage[dvips]{graphicx}
\usepackage{amsmath,amssymb}
\usepackage{algorithm}
\usepackage{algorithmic}

\usepackage[dvips]{graphicx}
\usepackage{amsmath,amssymb}
\usepackage{algorithm}
\usepackage{algorithmic}
\usepackage{amsmath,amssymb,amsbsy,amsthm,setspace}

\let\epsilon\varepsilon
\let\phi\varphi

\let\epsilon\varepsilon

\newtheorem*{lemma*}{Lemma}
\newtheorem{theorem}{Theorem}
\newtheorem{lemma}{Lemma}

\newtheorem{definition}{Definition}

\begin{document}
\title{Information-Theoretical Analysis of Two Shannon's  Ciphers }
\date{}

\author{
\IEEEauthorblockN{Boris Ryabko}
\IEEEauthorblockA{Institute of Computational Technologies of Siberian Branch of \\ the Russian Academy of Science,\\
Novosibirsk State University \\ Novosibirsk, Russia \\ 
}
}

\maketitle

\begin{abstract}
We describe generalized running key ciphers and apply them for analysis of two Shannon's methods.
In particular, we suggest some estimation of the cipher equivocation and the probability of correct deciphering without key.
\end{abstract}

\textbf{keywords:}  
running-key cipher,  cipher equivocation, Shannon cipher.

\section{Introduction}

We consider a classical problem of transmitting secret messages from a sender (Alice) to a
receiver (Bob) via an open channel which can be accessed by an adversary (Eve). It is assumed
that Alice and Bob (but not Eve) share a key, 
 which is a word in a certain alphabet.
Before transmitting a message to Bob, Alice encrypts it, and Bob, having received an encrypted
message (ciphertext), decrypts it to recover the plaintext.

We consider the so-called running-key ciphers where the plaintext
$X^1_1 ... X^1_t$, the key sequence
 $X^2_1 ... X^2_t$, and the ciphertext
 $Z_1 ...  Z_t$, 
are sequences of letters from 
 the same alphabet
$A = \{0,1, ... , n - 1\}$, where   $n \ge 2$. 
We assume that enciphering and deciphering 
 are given by the rules
$
Z_i = c(X^1_i,X^2_i), \, $ $ \, i = 1, ... , t, \,
X^1_i = d(Z_i,X^2_i), \, $ $ \, i = 1, ... , t
$, so that  $d(e(X^1_i,X^2_i), X^2_i) = X^1_i$.
C.Shannon in \cite{S} notes the following: 

{\it 
``The running key cipher can be easily improved to lead to ciphering systems which could not be solved without the key.
If one uses in place of one English text, about $d$ different texts as key, adding them all to the message, a sufficient
amount of key has been introduced to produce a high positive equivocation.
Another method would be to use, say, every 10th letter of the text as key. The intermediate letters are
omitted and cannot be used at any other point of the message. This has much the same effect,
since these spaced letters are nearly independent.''
}

More formally, we can introduce the first cipher from Shannon's description above 
 as follows: there are  $s$ sources 
 $X^1, X^2, ... , X^s$, $s \ge 2$,
 and any $X^i$  generates letters from the alphabet $A = \{0,1, ... , n - 1\}$.
 Suppose that   $X^1$ is the plaintext, whereas  $X^2, ... , X^s$ are key sequences
 The ciphertext $Z$ is obtained as follows   
\begin{equation}\label{vs} 
Z_i = ( ( ... (X^1_i + X^2_i)\mod n +   X^3_i)\mod n )      + ... +  X^s_i)\mod n \,. 
\end{equation}
The deciphering is obvious. 
In this report we perform information-theoretical analysis of both Shannon's ciphers.
\section{Estimations of the equivocation} 
For stationary ergodic processes $W$,$V$ and   $t \ge 1$ the $t$-order entropy is given by: 
$$
 h_t(W) = - t^{-1} \sum_{u \in A^{t-1} } P_W(u) \sum_{v \in A } P_W(v|u) \log_2 P_W(v|u)
\, . $$ 
The conditional entropy is   $ h_t(W/V) =   h_t(W,V) - h_t(V)$, see \cite{Cover:06}. 
In  \cite{S} Shannon called  $h_t(X^1/Z)$   the cipher equivocation and showed that the larger the equivocation, the better the cipher. 
Unfortunately, there are many cases where a direct calculation of 
the  equivocation is impossible and an  estimation is needed.
The following lemma can be used for this purpose. 
\begin{lemma}
 Let  $X^1, X^2, $ $ ... ,$ $ X^s$, $s \ge 2$, be  $s$-dimensional stationary ergodic source  and  $X^1, X^2, $ $ ... ,$ $ X^s$ be independent.  
If the cipher  (\ref{vs}) is applied, then
for any $t \ge 1$  
\begin{equation}\label{l1}
 h_t(X^1/Z) +  h_t(X^2/Z) + ... + h_t(X^{s-1}/Z) \ge $$ $$h_t(X^1)  +  h_t(X^2) + ... + h_t(X^s ) - \log_2 n
\end{equation}
and 
\begin{equation}\label{l2}
\frac{s-1}{s} ( h_t(X^1/Z) +  h_t(X^2/Z) + ... + h_t(X^{s}/Z)) \ge $$ $$h_t(X^1)  +  h_t(X^2) + ... + h_t(X^s ) - \log_2 n
\end{equation}
 \end{lemma}
 The proof of this lemma is given in the Appendix.
 \begin{definition} 
 Denote
 \begin{equation}\label{la}
  \Lambda_t = \frac{1}{s-1} ( h_t(X^1)  +  h_t(X^2) + ... + h_t(X^s )- \log_2 n ) \, .
 \end{equation} \end{definition}
 Note that, if $X^i, i = 1, ..., s,$ have the same 
 distribution, then  
 \begin{equation}\label{l3}
  \Lambda_t =  \frac{1}{s-1} (s \,  h_t(X^1) - \log_2 n  ) \, .
 \end{equation}

The following definition is due 
 to Lu Shyue-Ching \cite{lu}
   \begin{definition} 
 Let $M = M(Z_1 ... Z_t)$ $ = X^{1 *}_1 ... X^{1 *}_t$ be a certain function over $A^t$. Define
 $p_b$ $  = (1/t) $ $ \sum_{i=1}^t P(  X^{1 *}_i = X^1_i )$.
  \end{definition}
Note that, if $M $ is a method of deciphering $Z_1 ... Z_t$ without the key, then  $p_b$  is the
 average probability of  deciphering 
a single letter correctly.
 Obviously, the smaller  $p_b$, the better the cipher.
 Now we investigate the two methods of Shannon  described above.
             
\begin{theorem}\label{T1}
Let $X^1,$ $ X^2,$ $ ... ,$ $ X^s$, $s \ge 2$, be  $s$-dimensional stationary ergodic process           and
 $X^1,$ $ X^2,$ $ ... ,$ $ X^s$ be independent. Suppose that the cipher    (\ref{vs}) is applied. 
Then
\begin{itemize}
 \item[i)] The following inequality for the average probability $p_b$ is valid 
$$ (1 - p_b) \log_2 (n-1) + h_1(p_b) \ge \Lambda_t  \, ,
$$ 
\item[ii)] for any  $\delta > 0, \varepsilon > 0 $ there exists  $t^*$ such that for $t > t^*$ there exists a set $\Psi$ 
of texts of length $t$
for which
 $(P(\Psi) ) > 1- \delta$, for any  $V_1 ... V_t $ $\in \Psi$, $U_1 ... U_t $ $\in \Psi$
$$
(1/t) (\log_2 P(V_1 ... V_t)  - \log_2 P(U_1 ... U_t) 
 ) < \varepsilon 
$$ and 
$ \lim \inf_{t \rightarrow \infty } \frac{ 1}{t} \log_2 | \Psi |  \ge \Lambda_t $.
\end{itemize}
\end{theorem}
The proof of the first statement can be obtained by a method from \cite{lu}, whereas the second statement can be derived from \cite{rya}.

This theorem shows that if $\Lambda_t $ is large then the probability to find letters of the plaintext without the key must be small.
Besides, the second statement shows that Eve has a large set of possible plaintext whose probabilities are close and the total 
probability is closed to 1.

\section{Shannon ciphers}
Let us come back to the Shannon's methods  described above.
In \cite{se} Shannon estimated the entropy of printed English.
In particular, he showed that the entropy of the first order 
is approximately 
 4.14 for texts without 
spaces and 4.03 for texts with spaces.
He also estimated the limit entropy
to be  approximately 1 bit. 

Now we can investigate the first Shannon's cipher. 
 He suggested to use a sum of $d$ English texts as a key, i.e. use (\ref{vs}) with 
$s = d+1$, and where $X^i, i + 1, ... , s,$ is a text in printed English. 
Having taken into account that all $X^i$ are identically distributed,
 we immediately obtain from (\ref{l3}) the following:
 $$  \Lambda_t =  \frac{1}{s-1} (s \,  h_t(X^1) - \log_2 n  ) \, $$
 Taking into account that $ h_t(X^1) \approx  h_\infty(X^1) \approx  1$  and the estimation
  $\log_2 26 \approx 4.7$, 
we obtain the following approximation 
 $$  \Lambda_t =  \frac{1}{s-1} (s \,  h_t(X^1) - \log_2 n  ) \, = \,  \frac{1}{s-1} (s  - 4.7 ) \, . $$
 So, we can see that $ \Lambda_t $ is positive if $s -1 \ge 4$. Moreover, Theorem 1
 shows that  the first cipher of Shannon
 cannot be deciphered without the key if $d \ge 4$   different
texts are added  to the message (i.e., used as a key).

Let us consider the second cipher of Shannon. Here $s=2$,  a sequence $X^1$ is a  text in printed English and letters of $X^2$ are
generated independently with probabilities equal to the frequencies of occurrence of letters in English. 
From (\ref{vs}) we obtain
$$
\Lambda_t = ( h_t(X^1)  +  h_t(X^2) - \log_2 n ) \, .
$$
Having taken into account that  $ h_t(X^1) \approx  h_\infty(X^1) \approx  1$ , $ h_t(X^2) \approx 4.14$ (see \cite{se}) 
and $\log_2 26 \approx 4.7$, we can see
that $
\Lambda_t = 1 + 4.14  - 4.7 = 0.44$. So, $
\Lambda_t $ is positive and Theorem 1
 shows that  the first cipher of Shannon
 cannot be deciphered without key.

\section{Appendix}
\begin{proof}[Proof of the Lemma] 
The following chain of equalities and inequalities is valid:
$$  h_t(X^1) +  h_t(X^2) + ... +  h_t(X^s) =  h_t(X^1, X^2, ... , X^s) $$
$$ = h_t(X^1, X^2, ... , X^s, Z) = h_t(Z) + h_t(X^1, X^2, ... , X^s/ Z) $$
$$ =   h_t(Z) + h_t(X^1/ Z) + h_t( X^2/X^1, Z) + h_t( X^3/X^1, X^2, Z) + $$ $$ ... + h_t(X^s/X^1, X^2, ... , X^{s-1}, Z) $$
$$ =   h_t(Z) + h_t(X^1/ Z) + h_t( X^2/X^1, Z) + h_t( X^3/X^1, X^2, Z) + $$ $$ ... + h_t(X^{s-1}/X^1, X^2, ... , X^{s-2}, Z) $$
$$ \le  h_t(Z) + h_t(X^1/ Z) + h_t( X^2/Z) + h_t( X^3/Z) +   ... + h_t(X^{s-1}/ Z) .$$
The proof is based on well-known properties of the Shannon entropy which can be found, for example, in \cite{Cover:06}.
More precisely,  the first equation follows from the independence of   $X^1, X^2, ... , X^s$, whereas the second  equation is valid because   $Z$ is 
a function of $X^1, X^2, ... , X^s$, see (\ref{vs}). The third equation is a well-known property of the entropy. Having taken into account  
that $X^s$ is determined if $ X^2, ... , X^{s-1}, Z$ are known, we obtain the last equation. The inequality also follows from the properties of the 
Shannon entropy \cite{Cover:06}. 
Thus, \begin{equation}\label{in}
h_t(X^1) +  h_t(X^2) + ... +  h_t(X^s) \le $$ $$  h_t(Z) + h_t(X^1/ Z) + h_t( X^2/Z) + h_t( X^3/Z) +   ... + h_t(X^{s-1}/ Z) .  \end{equation}
Taking into account that for any process $U$ over alphabet $A = \{0, ... , n-1 \} $ $$ h_t(Z) \le \log_2 n \, , $$ we obtain (\ref{l1}) from (\ref{in}).
In order to prove (\ref{l2}) we note that analogously  to (\ref{in}), we can obtain the following:
$$ h_t(X^1) +  h_t(X^2) + ... +  h_t(X^s) \le $$ $$ \sum_{i=1}^{j-1} h_t(X^i/ Z) + \sum_{i=j+1}^{s} h_t(X^i/ Z)  $$  
for any $ 1 \le j \le s$.
From this inequality we obtain  (\ref{l2}).
\end{proof}

\section*{Acknowledgment}
This research  was supported  by  Russian Foundation for Basic Research
(grant no. 15-29-07932).

\end{document}